\documentclass[twocolumn,amsmath,amssymb,superscriptaddress,prb,showpacs]{revtex4}

\usepackage{graphicx}
\usepackage[euler]{textgreek}

\usepackage[breaklinks]{hyperref}
\hypersetup{
    bookmarks=false,
    pdfstartview={FitH},
    colorlinks=true,
    linkcolor=blue,
    citecolor=blue,
    urlcolor=blue
}

\begin{document}

\title{Direct Observation of Decoupled Structural and Electronic Transitions
and an Ambient Pressure Monoclinic-Like Metallic Phase of VO$_2$}

\author{J.~Laverock}
\affiliation{Department of Physics, Boston University, 590 Commonwealth Avenue,
Boston, MA 02215, USA}

\author{S.~Kittiwatanakul}
\affiliation{Department of Physics, University of Virginia,
Charlottesville, VA 22904, USA}

\author{A.~A.~Zakharov}
\affiliation{MAX-lab, Lund University, SE-221 00 Lund, Sweden}

\author{Y.~R.~Niu}
\affiliation{MAX-lab, Lund University, SE-221 00 Lund, Sweden}

\author{B.~Chen}
\affiliation{Department of Physics, Boston University, 590 Commonwealth Avenue,
Boston, MA 02215, USA}

\author{S.~A.~Wolf}
\affiliation{Department of Physics, University of Virginia,
Charlottesville, VA 22904, USA}
\affiliation{Department of Materials Science and Engineering,
University of Virginia, Charlottesville, VA 22904, USA}

\author{J.~W.~Lu}
\affiliation{Department of Materials Science and Engineering,
University of Virginia, Charlottesville, VA 22904, USA}

\author{K.~E.~Smith}
\affiliation{Department of Physics, Boston University, 590 Commonwealth Avenue,
Boston, MA 02215, USA}
\affiliation{School of Chemical Sciences and The MacDiarmid Institute for
Advanced Materials and Nanotechnology, The University of Auckland, Private
Bag 92019, Auckland 1142, New Zealand}


\begin{abstract}
We report
the simultaneous measurement of the structural and electronic components of the
metal-insulator transition of VO$_2$ using electron and photoelectron
spectroscopies and microscopies. We show that these evolve over different
temperature scales, and are
separated by an unusual monoclinic-like metallic phase.
Our results provide conclusive evidence that the new
monoclinic-like metallic phase, recently identified in high-pressure and
nonequilibrium measurements, is accessible in the thermodynamic transition at
ambient pressure, and we discuss the implications of these observations on the
nature of the MIT in VO$_2$.
\end{abstract}

\maketitle

The metal-insulator transition (MIT) of VO$_2$ is one of the most intensively
studied examples of its kind, and yet it continues to surprise and inform us:
some recent examples include the observation of its solid-state triple-point,
which is remarkably found to lie at the ambient pressure transition
temperature,\cite{park2013} and the peculiar nanosized striped topographical
pattern that has been found in strained VO$_2$ films.\cite{liu2013,liu2014}
Moreover, the phase transition itself faces renewed questions as to its
origin and mechanism following the discovery at high pressure, and in
nonequilibrium experiments, of a metallic state of monoclinic symmetry,
\cite{arcangeletti2007,hsieh2014,kim2006etc} which beforehand
had universally been the reserve of the insulating state in experiments. Very
recently, the decoupling of the structural and electronic phase transitions
has been confirmed in the related compound, V$_2$O$_3$.\cite{ding2014}
In part, the widespread interest that VO$_2$ has attracted is owed to the
accessibility of its sharp,\cite{morin1959etc}
ultrafast\cite{cavalleri2001}
transition, occurring in the bulk at 65~$^{\circ}$C at ambient pressures,
coupled with the rich tunability of its properties with alloying and
strain\cite{pouget1974,villeneuve1972,pouget1975} and flexibility in
fabrication\cite{muraoka2002etc} that make it a promising candidate
for device application.\cite{appavoo2014}

In the bulk, the MIT of VO$_2$ is accompanied by a large structural
distortion that has added to the difficulties in unraveling its origins. The
high temperature metallic phase resides in the tetragonal rutile structure
(isostructural with TiO$_2$). Below the first-order transition temperature,
V-V dimers form, accompanied by the twisting of the VO$_6$ octahedra, as
the structure is distorted into the insulating monoclinic $M_1$ phase. A second
insulating monoclinic structure ($M_2$), in which one-half of the V atoms
dimerize, is accessible through Cr doping\cite{pouget1974} and
strain.\cite{pouget1975} On
the one hand, the dimerization has been considered a hallmark of the
Peierls transition, in which the rearrangement of the lattice plays the
key role. On the other hand, several experiments have made it clear that
electron-electron correlations cannot be ignored,\cite{zylbersztejn1975}
and should be considered on at least an equal footing.\cite{haverkort2005}

We report the direct observation of the structural and electronic
components of the transition in strained VO$_2$ by {\em
simultaneously} combining powerful spatial and energy resolved
probes of the crystal and electronic structure. We further show that the
recently-observed monoclinic metallic
phase is accessible in the ground state of strained VO$_2$
at ambient temperatures and pressures.

\begin{figure*}[t!]
\begin{center}
\includegraphics[width=0.8\linewidth,clip]{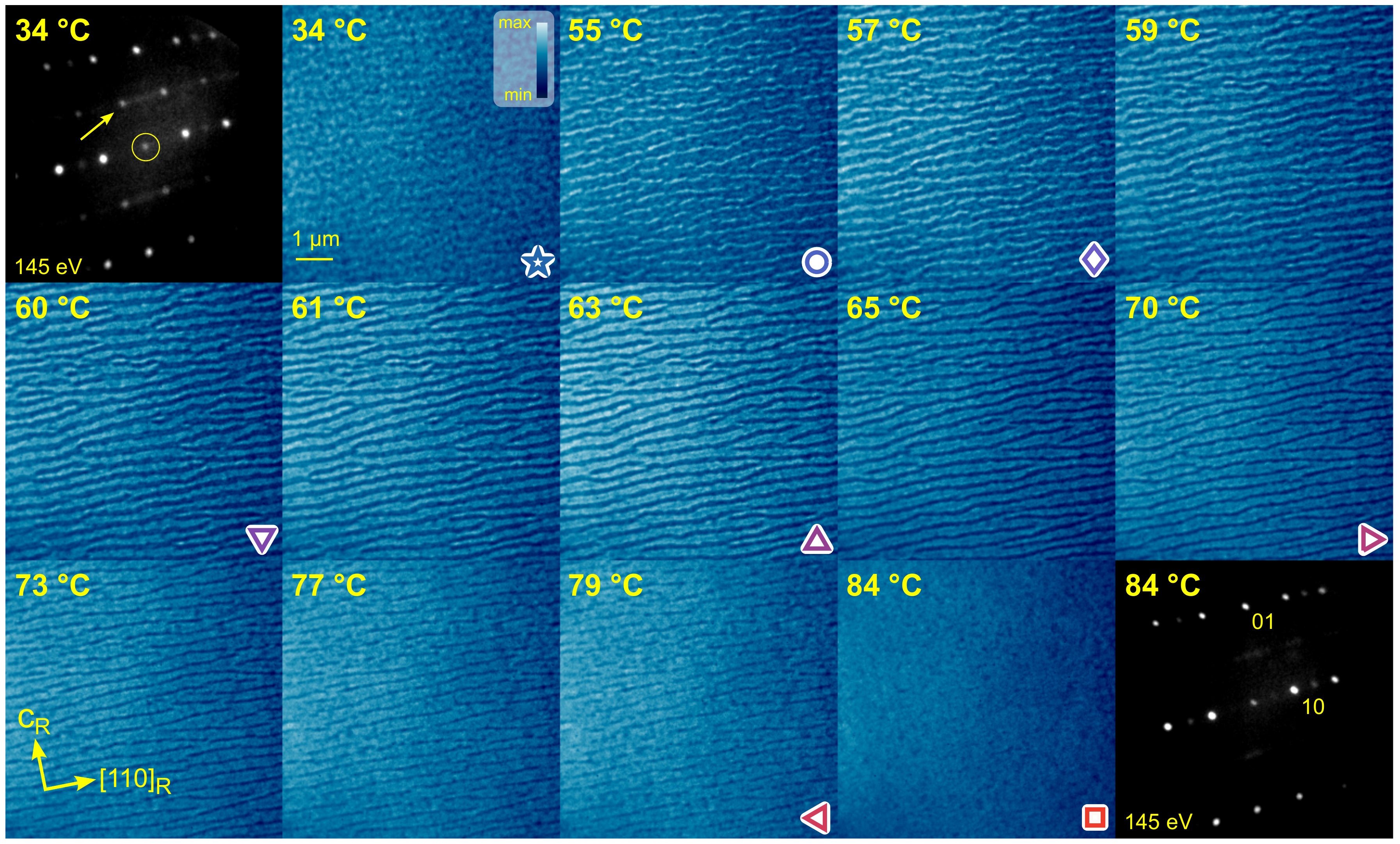}
\end{center}
\vspace*{-0.2in}
\caption{\label{f:leem} (Color online) Sequence of LEEM images of VO$_2$(110)
recorded at 10~eV. The symbols
correspond to the simultaneous acquisition of a PES spectrum [see
Fig.~\ref{f:pes}(a)]. LEED patterns at both endpoints are also shown;
the $(0,0)$ spot is circled, and the arrow indicates
the extra reflections in the low-symmetry monoclinic phase.}
\end{figure*}

\begin{figure*}[t!]
\begin{center}
\includegraphics[width=1.0\linewidth,clip]{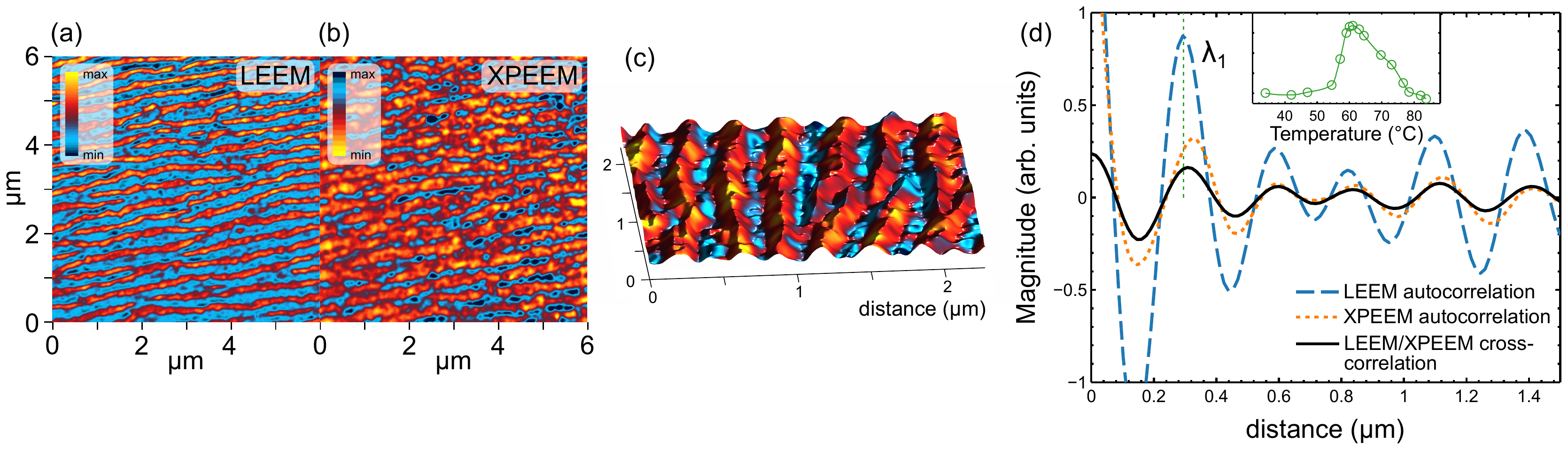}
\end{center}
\vspace*{-0.2in}
\caption{\label{f:xpleem} (Color online) Correspondence between LEEM and XPEEM
images, recorded at
56~$^{\circ}$C. (a)~LEEM image at 10~eV, and (b)~XPEEM image of
the spatially resolved work function
at a photon energy of 50~eV.
In (c), the XPEEM intensity in (b) is mapped as false color onto the surface
of the LEEM intensity in (a).
(d)~Autocorrelation of the LEEM and XPEEM
images perpendicular to the stripe direction, shown alongside
their cross-correlation. $\lambda_1$ indicates the location of the
first peak in the autocorrelation of the LEEM intensity, and the
temperature dependence of its magnitude is shown in the inset.}
\end{figure*}

High quality 110~nm thin films of VO$_2$ (r.m.s.\ roughness of 0.17~nm)
were grown on (110)-oriented substrates of rutile TiO$_2$, as described
previously,\cite{west2008betc} and are hereafter referred to as
VO$_2$(110). Electron
and photoelectron spectroscopy measurements were performed
at the SPELEEM endstation of Beamline I311, MAX-lab (Lund, Sweden), and the
samples were prepared for ultrahigh vacuum measurements as described
previously.\cite{laverock2012,laverock2012b}
Low-energy electron diffraction (LEED) patterns were collected from a
5~{\textmugreek}m diameter region of the sample, with the electron emission
current restricted to less than 10~nA to minimize radiation damage.
Low-energy electron microscopy (LEEM) images
were corrected for the non-uniform detector efficiency and background before
histogramming; raw images are shown in Fig.~\ref{f:leem}, and following
correction in Fig.~\ref{f:xpleem}(a).

Bright-field LEEM images of the surface of
VO$_2$(110)
are shown in Fig.~\ref{f:leem}, recorded at an electron energy of
10~eV across the MIT. In this regime, the contrast of LEEM originates from
differences in the $(0,0)$ diffraction intensity within the first few atomic
layers,\cite{bauer1994}
and probes the local crystal structure of VO$_2$.
LEED patterns (Fig.~\ref{f:leem})
were also recorded both above and below the transition and confirm the
evolution in the crystal structure across the MIT.
Above the transition, the LEED
pattern at 145~eV resembles the familiar rutile pattern,\cite{goering1997}
but at 34~$^{\circ}$C additional superstructure spots
are visible due to the lower symmetry of the
$(110)_{\rm R}$ monoclinic surface. These measurements also revealed a strong
and reproducible sensitivity of the sample surface to both electron and photon
irradiation. We emphasize that in all subsequent measurements care was taken to
minimize the radiation damage. In particular, exposure to the photon flux
was limited to less than 1~min per spectrum, with LEED patterns checked
before and after the measurements.

In the monoclinic insulating phase at 34~$^{\circ}$C, the LEEM image exhibits
a slightly
rough appearance, anticipating the emergence of the rutile stripes. At
55~$^{\circ}$C, thin stripes of higher LEEM intensity become clearly visible,
oriented along the rutile $[110]_{\rm R}$ crystal direction, in agreement
with the previous atomic force microscopy (AFM) and
infrared images.\cite{liu2013} These (bright)
domains represent regions of the sample with a different crystallographic
structure to the insulating monoclinic phase, and can be associated with the
transition to the rutile (metallic) phase. As the temperature is increased,
the rutile stripes rapidly grow in size until $\approx 65$~$^{\circ}$C, after
which there is a more gradual growth until the stripe structure disappears
at 82~$^{\circ}$C and the system is fully rutile.

The phase separation between monoclinic and rutile structures we
observe in LEEM is reinforced by x-ray photoemission electron microscopy (XPEEM)
measurements of the onset of
secondary electrons, which provides direct spatial information on the
work function of the surface.
From photoemission spectroscopy (PES)
measurements of the low and high temperature phases, we
find that the work function of rutile VO$_2$(110) is 0.12~eV larger than
monoclinic VO$_2$(110), in good agreement with recent Kelvin force probe
measurements,\cite{ko2011} although opposite to previous PES results on VO$_2$
``nano-bundles''.\cite{yin2011} Since the work function is a property of
the material surface, and in particular the packing density, it offers
an alternative (electronic) perspective to LEEM of the lattice structure.
Figures~\ref{f:xpleem}(a) and \ref{f:xpleem}(b) show LEEM and XPEEM
images respectively of the mixed phase
of VO$_2$(110) at 56~$^{\circ}$C on the same part of the sample. Since the
work function of rutile VO$_2$ is larger than that of monoclinic VO$_2$,
the XPEEM contrast is inverted with respect to LEEM, and for clarity the
color scale in Fig.~\ref{f:xpleem}(b) has been reversed compared to (a). The
similarity in the structure between the two images is clear: in addition to
the rutile stripes observed in XPEEM, it is possible to identify the same
forks in this pattern in both images, e.g.~at
$(x,y) = (3.3, 2.0)$~{\textmugreek}m
and $(1.0,2.0)$~{\textmugreek}m. In Fig.~\ref{f:xpleem}(c), the XPEEM
intensity has been mapped as color onto the LEEM intensity, which is shown as
a 2D surface, providing a direct visualization of the spatial correlation
between the two probes of the lattice structure: diffraction and work function
(electronic).

In Fig.~\ref{f:xpleem}(d), the autocorrelation of the LEEM (i.e.~$I_{\rm
LEEM} \star I_{\rm LEEM}$) and XPEEM images is shown along the $c_{\rm R}$
direction (perpendicular to the rutile stripes). Both functions exhibit
a strong, damped cosinusoidal form of the same periodicity, typical of
a regularly ordered pattern. Also shown in Fig.~\ref{f:xpleem}(d) is the
(inverted) cross-correlation between LEEM and XPEEM, $I_{\rm LEEM} \star
I_{\rm XPEEM}$, which has the same form as the autocorrelation curves,
and persists over remarkably long lengthscales. The first peak in the
autocorrelation, labeled $\lambda_1$ in Fig.~\ref{f:xpleem}(d), corresponds to
the mean distance between stripes, and is $\approx 300$~nm. The
temperature dependence of the magnitude of this peak is shown in the inset
to Fig.~\ref{f:xpleem}(d). The rapid rise above 55~$^{\circ}$C corresponds
to the stripe formation, and the shoulder at 70~$^{\circ}$C reflects their
more gradual growth towards the end of the transition. The strong quantitative
correlation between LEEM and XPEEM illustrates the structural
nature of the stripe pattern, in agreement with the topographic rumpling of
the surface previously observed via AFM.\cite{liu2013}

\begin{figure*}[t!]
\begin{center}
\includegraphics[width=1.0\linewidth,clip]{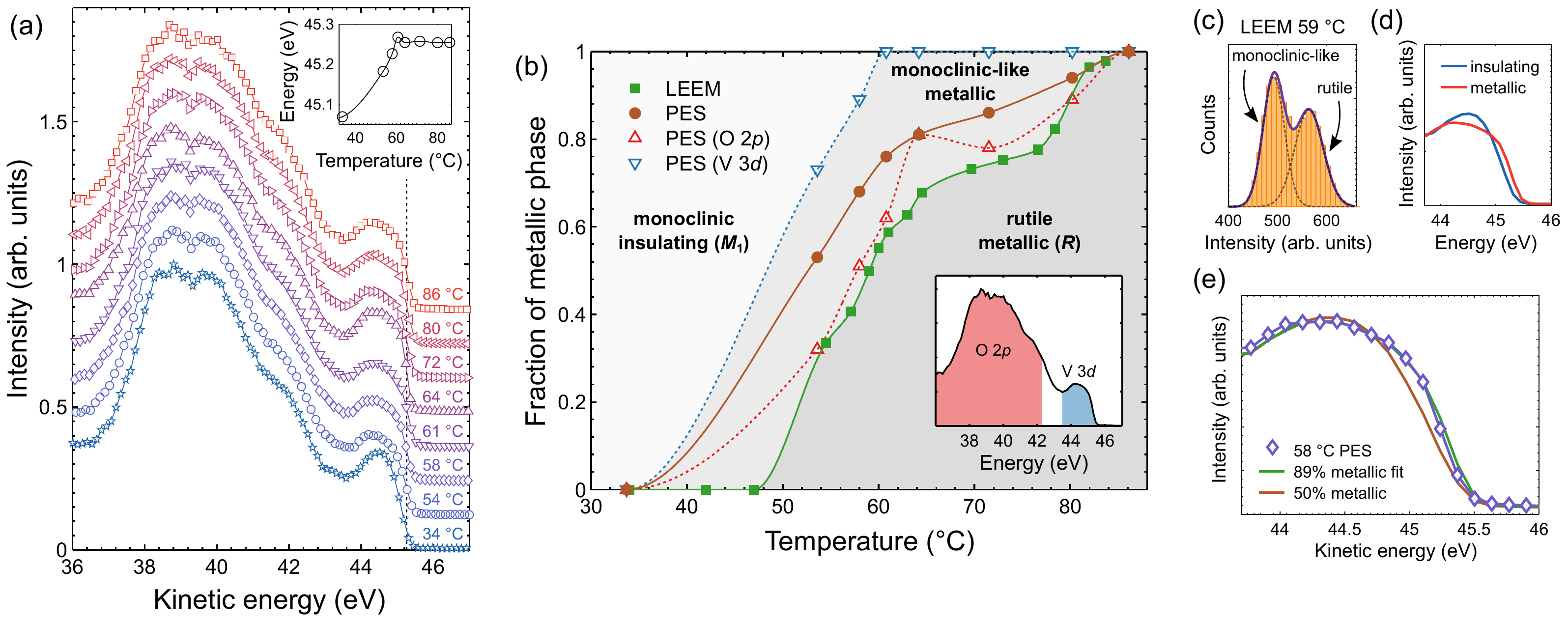}
\end{center}
\vspace*{-0.2in}
\caption{\label{f:pes} (Color online)
(a) PES measurements of VO$_2$(110) across the MIT
recorded on
the same part of the sample as the LEEM images of Fig.~\ref{f:leem}. The
upper inset illustrates the evolution of the leading edge of the spectra
with temperature.
(b)~The fraction of the metallic phase extracted from LEEM and
PES (the lines are guides for the
eye). The inset shows the regions-of-interest used for the separate analysis
of O~$2p$ and V~$3d$ states. (c)~Histogram of the LEEM intensity at
59~$^{\circ}$C. (d)~Magnified view of the V~$3d$ states in the insulating and
metallic phases. (e)~Results of fitting the insulating and metallic end-points
to the 58~$^{\circ}$C PES data in the V~$3d$ region (89\% metallic fraction).
The spectrum corresponding to the LEEM results (50\% metallic fraction) is
also shown for comparison.}
\end{figure*}

We now turn to PES to
directly explore the electronic behavior of the stripes. At selected
temperatures through
the MIT, PES spectra were simultaneously
recorded (within 1~min of the corresponding LEEM image) at a photon energy of
50~eV, and are shown in Fig.~\ref{f:pes}(a).
The PES spectra
are composed of O $2p$ states between 36 and 43~eV, and V $3d$ states near
the Fermi level above 43~eV, and are qualitatively similar to previous
PES measurements.\cite{okazaki2004etc,laverock2012b,koethe2006etc}
Below the MIT
(bottom spectrum), the V $3d$ states are relatively narrow, whereas above
the transition the transfer of spectral weight into quasiparticle states at
the Fermi level (higher kinetic energies) indicates the formation of the
metallic phase.  The top inset to Fig.~\ref{f:pes}(a) shows the evolution
in the leading edge of the V $3d$ states with temperature, determined by
locating the extrema in the derivative of the PES. This quantity is found
to shift to higher energies by 0.19~eV over the measured temperature range,
and is in good agreement with high-resolution dichroic PES measurements.
\cite{laverock2012b}

The LEEM and PES results are compared in Fig.~\ref{f:pes}(b) by analyzing the
fractions of their constituent components with temperature. At the energies
employed in this study, both LEEM and PES (and, indeed, LEED) have
similar depth sensitivities (of $\lesssim 1$~nm), meaning our results probe
essentially the same physical volume of the sample. Histograms were
constructed of the intensity of each LEEM image in Fig.~\ref{f:leem}, which
were fitted to either one or two Gaussian components (below
47~$^{\circ}$C only a single component
could be resolved). For the other images, the fraction of the brighter component
was associated with the rutile phase [as shown in Fig.~\ref{f:pes}(c)],
and the results are shown by the squares
in Fig.~\ref{f:pes}(b). Correspondingly, the fraction of the metallic phase has
been estimated from the PES data by assuming the end-points (at 34~$^{\circ}$C
and 86~$^{\circ}$C) are representative of each phase. The intermediate spectra
have been fitted to a linear combination of these two end-points (see
Ref.~\onlinecite{laverock2012b} for an example of this procedure), and the
results are shown by the circles in Fig.~\ref{f:pes}(b). We emphasize that we do
not find evidence of a third component in either LEEM or PES analyses, although
we cannot rule out such a phase below our detection level. The plateau in the
progression of the MIT discussed above is clearly evident in both LEEM and
PES data between 65~$^{\circ}$C and 75~$^{\circ}$C in Fig.~\ref{f:pes}(b),
and may be related to the interface energy of the stripe domains.

The good qualitative agreement in Fig.~\ref{f:pes}(b) is mitigated slightly
by the higher fraction of the metallic phase extracted from the PES data.
To gain additional insight, we have repeated this procedure focusing separately
on the O~$2p$ and V~$3d$ states by restricting the regions-of-interest of
the fit [shown in the inset to Fig.~\ref{f:pes}(b)]. Remarkably,
the metallic fraction determined from the O~$2p$ states alone
is found to closely follow that determined from LEEM. Since the
O~$2p$ states are most sensitive to their bonding environment (i.e.~the
structure of the material), this provides a satisfying quantitative link
between LEEM and PES. On the other hand, the metallic fraction determined
from the V~$3d$ states shows quite different behavior:
it initially rapidly rises before saturating near 60~$^{\circ}$C, strongly
reminiscent of the temperature evolution of the leading edge of the PES shown
in Fig.~\ref{f:pes}(a). A specific example is shown in Fig.~\ref{f:pes}(c-e),
in which the histogram of the LEEM intensity at 59~$^{\circ}$C~(c) clearly
shows two components of approximately the same area (the metallic fraction is
determined to be 50\%). However, the V~$3d$
spectrum at this temperature~(e) cannot be described as a 50:50 average
of the insulating and metallic end-points~(d). Instead, it is
well approximated if a metallic fraction of 89\% is assumed~(e).
A previous nanoscale imaging study, combining structural (diffraction) and
electronic (infrared scattering) probes, also noticed differences in the
progression of the structural and electronic components in
VO$_2$ thin films,\cite{qazilbash2011} although we do not observe the
non-monotonic evolution in the structure reported by those authors.

Taken together, these results reveal a separation of the temperature
scales of the structural and electronic transitions in VO$_2$(110),
the former of which is not complete until $\sim 84$~$^{\circ}$C
whereas the latter becomes fully metallic (within our precision) at
$\approx 61$~$^{\circ}$C. At intermediate temperatures
(60~--~80~$^{\circ}$C) VO$_2$(110) consists of a mixture of rutile metallic
and monoclinic-like metallic phases, where we use the qualifier ``like''
to indicate the LEEM intensity and O~$2p$ photoelectrons closely resemble
the monoclinic $M_1$ phase. At lower temperatures (50~--~60~$^{\circ}$C) all
three phases
are in equilibrium. Such monoclinic metallic states in VO$_2$ have previously
been identified under high pressure\cite{arcangeletti2007,hsieh2014} and in
out-of-equilibrium measurements (e.g.~photoexcited pump-probe, charge-doped
or voltage-driven
experiments).\cite{kim2006etc} In our strained VO$_2$(110)
film, the $a$ and $b$ lattice parameters (determined from x-ray diffraction)
correspond to
a compressive strain along the [110] axis of $\sim 2$\%, which corresponds to
approximately 12~GPa of hydrostatic pressure, very close to the onset of the
monoclinic metallic phase in high-pressure measurements.\cite{mitrano2012}
Our results reveal that this unusual new phase is accessible {\em in the ground
state} at ambient temperatures and pressures in epitaxially strained VO$_2$.

The full structural and electronic details of monoclinic-like metallic phase(s)
in VO$_2$ have yet to be experimentally reported, although there are already
several hints. For example, detailed structural measurements of bulk VO$_2$,
accompanied by band structure calculations, suggest the melting of the
V-V dimers may stabilize the metallic state before the tetragonal symmetry
is adopted,\cite{yao2010} a picture that is supported by first-principles
calculations of the photoinduced transition.\cite{yuan2013etc}
At high pressure, Raman measurements indicate a rearrangement of the
V-V dimers.\cite{arcangeletti2007}

Given the available information, it is possible that the monoclinic-like
metallic phase that we observe develops due to the spontaneous breaking,
or substantial weakening, of the V-V bond, while the system remains monoclinic,
and this phase is stabilized by the high in-plane effective pressure due to
the substrate clamping. Alternatively, if the V-V dimers remain
strongly paired in this phase, electron-electron correlations may drive
the transition. It is therefore likely that the structural details in this
phase are capable of distinguishing the dominant interaction that
drives the MIT.
Ultimately, given the possible proximity of VO$_2$ to a
conventional Mott-Hubbard transition,\cite{lazarovits2010,laverock2012b}
coupled with the several structural instabilities accessible through pressure
and chemical doping,\cite{goodenough1973,pouget1974,park2013} it may turn
out that the photoinduced and thermodynamic monoclinic metallic phases
differ. Future planned measurements are required
to clarify the fate of the V-V dimers
in the monoclinic-like metallic phase, which may finally hold the key to
a deeper understanding of the microscopic mechanism of the MIT in VO$_2$.

{\bf Acknowledgements}. The Boston University program is supported in part by
the Department of Energy under Grant No.\ DE-FG02-98ER45680. S.K., J.W.L.\ and
S.A.W.\ are
thankful to the financial support from the Army Research Office through MURI
grant No.\ W911-NF-09-1-0398.


\begin{thebibliography}{99}

\bibitem{park2013}
J.\ H.\ Park, J.\ M.\ Coy, T.\ S.\ Kasirga, C.\ Huang, Z.\ Fei, S.\ Hunter and
D.\ H.\ Cobden,
\href{http://dx.doi.org/10.1038/nature12425}
{Nature (London) {\bf 500}, 431 (2013)}.

\bibitem{liu2013}
M.\ K.\ Liu, M.\ Wagner, E.\ Abreu, S.\ Kittiwatanakul, A.\ McLeod, Z.\ Fei,
M.\ Goldflam, S.\ Dai, M.\ M.\ Fogler, J.\ Lu, S.\ A.\ Wolf, R.\ D.\ Averitt,
D.\ N.\ Basov,
\href{http://dx.doi.org/10.1103/PhysRevLett.111.096602}
{Phys.\ Rev.\ Lett.\ {\bf 111} 096602 (2013)}.

\bibitem{liu2014}
M.\ K.\ Liu, M.\ Wagner, J.\ Zhang, A.\ McLeod, S.\ Kittiwatanakul, Z.\ Fei,
E.\ Abreu, M.\ Goldflam, A.\ J.\ Sternbach, S.\ Dai, K.\ G.\ West, J.\ Lu,
S.\ A.\ Wolf, R.\ D.\ Averitt and D.\ N.\ Basov,
\href{http://dx.doi.org/10.1063/1.4869558}
{Appl.\ Phys.\ Lett.\ {\bf 104}, 121905 (2014)}.

\bibitem{arcangeletti2007}
E.\ Arcangeletti, L.\ Baldassarre, D.\ Di Castro, S.\ Lupi, L.\ Malavasi,
C.\ Marini, A.\ Perucchi and P.\ Postorino,
\href{http://dx.doi.org/10.1103/PhysRevLett.98.196406}
{Phys.\ Rev.\ Lett.\ {\bf 98}, 196406 (2007)}.

\bibitem{hsieh2014}
W.-P.\ Hsieh, M.\ Trigo, D.\ A.\ Reis, G.\ A.\ Artioli, L.\ Malavasi and
W.\ L.\ Mao,
\href{http://dx.doi.org/10.1063/1.4862197}
{Appl.\ Phys.\ Lett.\ {\bf 104}, 021917 (2014)}.

\bibitem{kim2006etc}
H.-T.\ Kim, Y.\ W.\ Lee, B.-J.\ Kim, B.-G.\ Chae, S.\ J.\ Yun,
K.-Y.\ Kang, K.-J.\ Han, K.-J.\ Yee and Y.-S.\ Lim,
\href{http://dx.doi.org/10.1103/PhysRevLett.97.266401}
{Phys.\ Rev.\ Lett.\ {\bf 97}, 266401 (2006)};
B.-J.\ Kim, Y.\ W.\ Lee, S.\ Choi, J.-W.\ Lim, S.\ J.\ Yun, H.-T.\ Kim,
T.-J.\ Shin and H.-S.\ Yun,
\href{http://dx.doi.org/10.1103/PhysRevB.77.235401}
{Phys.\ Rev.\ B {\bf 77}, 235401 (2008)};
Z.\ Tao, T.-R.\ T.\ Han, S.\ D.\ Mahanti, P.\ M.\ Duxbury, F.\ Yuan,
C.-Y.\ Ruan, K.\ Wang and J.\ Wu,
\href{http://dx.doi.org/10.1103/PhysRevLett.109.166406}
{Phys.\ Rev.\ Lett.\ {\bf 109}, 166406 (2012)};
T.\ L.\ Cocker, L.\ V.\ Titova, S.\ Fourmaux, G.\ Holloway, H.-C.\ Bandulet,
D.\ Brassard, J.-C.\ Kieffer, M.\ A.\ El Khakani and F.\ A.\ Hegmann,
\href{http://dx.doi.org/10.1103/PhysRevB.85.155120}
{Phys.\ Rev.\ B {\bf 85}, 155120 (2012)}.

\bibitem{ding2014}
Y.\ Ding, C.-C.\ Chen, Q.\ Zeng, H.-S.\ Kim, M.\ J.\ Han, M.\ Balasubramanian,
R.\ Gordon, F.\ Li, L.\ Bai, D.\ Popov, S.\ M.\ Heald, T.\ Gog, H.-K.\ Mao and
M.\ van Veenendaal,
\href{http://dx.doi.org/10.1103/PhysRevLett.112.056401}
{Phys.\ Rev.\ Lett.\ {\bf 112}, 056401 (2014)}.

\bibitem{morin1959etc}
F.\ J.\ Morin,
\href{http://dx.doi.org/10.1103/PhysRevLett.3.34}
{Phys.\ Rev.\ Lett.\ {\bf 3}, 34 (1959)};
N.\ F.\ Mott,
{\em Metal-Insulator Transitions}, Taylor \& Francis Ltd, London (1974).

\bibitem{cavalleri2001}
A.\ Cavalleri, Cs.\ T\'{o}th, C.\ W.\ Siders, J.\ A.\ Squier, F.\ R\'{a}ksi,
P.\ Forget and J.\ C.\ Kieffer,
\href{http://dx.doi.org/10.1103/PhysRevLett.87.237401}
{Phys.\ Rev.\ Lett.\ {\bf 87}, 237401 (2001)}.

\bibitem{pouget1974}
J.\ P.\ Pouget, H.\ Launois, T.\ M.\ Rice, P.\ Dernier, A.\ Gossard, G.\
Villeneuve and P.\ Hagenmuller,
\href{http://dx.doi.org/10.1103/PhysRevB.10.1801}
{Phys.\ Rev.\ B {\bf 10}, 1801 (1974)}.

\bibitem{villeneuve1972}
G.\ Villeneuve, A.\ Bordet, A.\ Casalot, J.\ P.\ Pouget, H.\ Launois and
P.\ Lederer,
\href{http://dx.doi.org/10.1016/S0022-3697(72)80494-3}
{J.\ Phys.\ Chem.\ Solids {\bf 33}, 1953 (1972)}.

\bibitem{pouget1975}
J.\ P.\ Pouget, H.\ Launois, J.\ P.\ D'Haenens, P.\ Merenda and T.\ M.\ Rice,
\href{http://dx.doi.org/10.1103/PhysRevLett.35.873}
{Phys.\ Rev.\ Lett.\ {\bf 35}, 873 (1975)}.

\bibitem{muraoka2002etc}
Y.\ Muraoka and Z.\ Hiroi,
\href{http://dx.doi.org/10.1063/1.1446215}
{Appl.\ Phys.\ Lett.\ {\bf 80}, 583 (2002)};
J.\ Cao, E.\ Ertekin, V.\ Srinivasan, W.\ Fan, S.\ Huang, H.\ Zheng,
J.\ W.\ L. Yim, D.\ R.\ Khanal, D.\ F.\ Ogletree, J.\ C.\ Grossman and J.\ Wu,
\href{http://dx.doi.org/10.1038/nnano.2009.266}
{Nature Nanotech.\ {\bf 4}, 732 (2009)}.

\bibitem{appavoo2014}
K.\ Appavoo, B.\ Wang, N.\ F.\ Brady, M.\ Seo, J.\ Nag, R.\ P.\ Prasankumar,
D.\ J.\ Hilton, S.\ T.\ Pantelides and R.\ F.\ Haglund, Jr.,
\href{http://dx.doi.org/10.1021/nl4044828}
{Nano Lett.\ {\bf 14}, 1127 (2014)}.

\bibitem{zylbersztejn1975}
A.\ Zylbersztejn and N.\ F.\ Mott,
\href{http://dx.doi.org/10.1103/PhysRevB.11.4383}
{Phys.\ Rev.\ B {\bf 11} 4383 (1975)}.

\bibitem{haverkort2005}
M.\ W.\ Haverkort, Z.\ Hu, A.\ Tanaka, W.\ Reichelt, S.\ V.\ Streltsov, M.\ A.\
Korotin, V.\ I.\ Anisimov, H.\ H.\ Hsieh, H.-J.\ Lin, C.\ T.\ Chen, D.\ I.\
Khomskii and L.\ H.\ Tjeng,
\href{http://dx.doi.org/10.1103/PhysRevLett.95.196404}
{Phys.\ Rev.\ Lett.\ {\bf 95}, 196404 (2005)}.

\bibitem{west2008betc}
K.\ G.\ West, J.\ W.\ Lu, J.\ Yu, D.\ Kirkwood, W.\ Chen, Y.\ H.\ Pei,
J.\ Claassen and S.\ A.\ Wolf,
\href{http://dx.doi.org/10.1116/1.2819268}
{J.\ Vac.\ Sci.\ Technol.\ A {\bf 26}, 133 (2008)};
S.\ Kittiwatanakul, J.\ Laverock, D.\ Newby, Jr., K.\ E.\ Smith,
S.\ A.\ Wolf and J.\ Lu,
\href{http://dx.doi.org/10.1063/1.4817174}
{J.\ Appl.\ Phys.\ {\bf 114}, 053703 (2013)}.

\bibitem{laverock2012}
J.\ Laverock, L.\ F.\ J.\ Piper, A.\ R.\ H.\ Preston, B.\ Chen, J.\ McNulty,
K.\ E.\ Smith, S.\ Kittiwatanakul, J.\ W.\ Lu, S.\ A.\ Wolf, P.-A.\ Glans and
J.-H.\ Guo,
\href{http://dx.doi.org/10.1103/PhysRevB.85.081104}
{Phys.\ Rev.\ B {\bf 85}, 081104(R) (2012)}.

\bibitem{laverock2012b}
J.\ Laverock, A.\ R.\ H.\ Preston, D.\ Newby, Jr., K.\ E.\ Smith, S.\ Sallis,
L.\ F.\ J.\ Piper, S.\ Kittiwatanakul, J.\ W.\ Lu, S.\ A.\ Wolf, M.\
Leandersson and T.\ Balasubramanian,
\href{http://dx.doi.org/10.1103/PhysRevB.86.195124}
{Phys.\ Rev.\ B {\bf 86}, 195124 (2012)}.

\bibitem{bauer1994}
E.\ Bauer,
\href{http://dx.doi.org/10.1088/0034-4885/57/9/002}
{Rep.\ Prog.\ Phys.\ {\bf 57}, 895 (1994)}.

\bibitem{goering1997}
E.\ Goering, M.\ Schramme, O.\ M\"{u}ller, R.\ Barth, H.\ Paulin, M.\ Klemm,
M.\ L.\ denBoer and S.\ Horn,
\href{http://dx.doi.org/10.1103/PhysRevB.55.4225}
{Phys.\ Rev.\ B {\bf 55}, 4225 (1997)}.

\bibitem{ko2011}
C.\ Ko, Z.\ Yang and S.\ Ramanathan,
\href{http://dx.doi.org/10.1021/am2006299}
{ACS Appl.\ Mater.\ Interfaces {\bf 3}, 3396 (2011)}.

\bibitem{yin2011}
H.\ Yin, M.\ Luo, K.\ Yu, Y.\ Gao, R.\ Huang, Z.\ Zhang, M.\ Zeng, C.\ Cao and
Z.\ Zhu,
\href{http://dx.doi.org/10.1021/am200291a}
{ACS Appl.\ Mater.\ Interfaces {\bf 3}, 2057 (2011)}.

\bibitem{okazaki2004etc}
K.\ Okazaki, H.\ Wadati, A.\ Fujimori, M.\ Onoda, Y.\ Muraoka and Z.\ Hiroi,
\href{http://dx.doi.org/10.1103/PhysRevB.69.165104}
{Phys.\ Rev.\ B {\bf 69}, 165104 (2004)};
K.\ Saeki, T.\ Wakita, Y.\ Muraoka, M.\ Hirai, T.\ Yokoya, R.\ Eguchi and
S.\ Shin,
\href{http://dx.doi.org/10.1103/PhysRevB.80.125406}
{Phys.\ Rev.\ B {\bf 80}, 125406 (2009)}.

\bibitem{koethe2006etc}
T.\ C.\ Koethe, Z.\ Hu, M.\ W.\ Haverkort, C.\ Sch\"{u}{\ss}ler-Langeheine,
F.\ Venturini, N.\ B.\ Brookes, O.\ Tjernberg, W.\ Reichelt, H.\ H.\ Hsieh,
H.-J.\ Lin, C.\ T.\ Chen and L.\ H.\ Tjeng,
\href{http://dx.doi.org/10.1103/PhysRevLett.97.116402}
{Phys.\ Rev.\ Lett.\ {\bf 97}, 116402 (2006)};
R.\ Eguchi, M.\ Taguchi, M.\ Matsunami, K.\ Horiba, K.\ Yamamoto, Y.\ Ishida,
A.\ Chainani, Y.\ Takata, M.\ Yabashi, D.\ Miwa, Y.\ Nishino, K.\ Tamasaku,
T.\ Ishikawa, Y.\ Senba, H.\ Ohashi, Y.\ Muraoka, Z.\ Hiroi and S.\ Shin,
\href{http://dx.doi.org/10.1103/PhysRevB.78.075115}
{Phys.\ Rev.\ B {\bf 78}, 075115 (2008)}.

\bibitem{qazilbash2011}
M.\ M.\ Qazilbash, A.\ Tripathi, A.\ A.\ Schafgans, B.-J.\ Kim, H.-T.\ Kim,
Z.\ Cai, M.\ V.\ Holt, J.\ M.\ Maser, F.\ Keilmann, O.\ G.\ Shpyrko and
D.\ N.\ Basov,
\href{http://dx.doi.org/10.1103/PhysRevB.83.165108}
{Phys.\ Rev.\ B {\bf 83}, 165108 (2011)}.

\bibitem{mitrano2012}
M.\ Mitrano, B.\ Maroni, C.\ Marini, M.\ Hanfland, B.\ Joseph, P.\ Postorino
and L.\ Malavasi,
\href{http://dx.doi.org/10.1103/PhysRevB.85.184108}
{Phys.\ Rev.\ B {\bf 85}, 184108 (2012)}.

\bibitem{yao2010}
T.\ Yao, X.\ Zhang, Z.\ Sun, S.\ Liu, Y.\ Huang, Y.\ Xie, C.\ Wu, X.\ Yuan,
W.\ Zhang, Z.\ Wu, G.\ Pan, F.\ Hu, L.\ Wu, Q.\ Liu and S.\ Wei,
\href{http://dx.doi.org/10.1103/PhysRevLett.105.226405}
{Phys.\ Rev.\ Lett.\ {\bf 105}, 226405 (2010)}.

\bibitem{yuan2013etc}
X.\ Yuan, W.\ Zhang and P.\ Zhang,
\href{http://dx.doi.org/10.1103/PhysRevB.88.035119}
{Phys.\ Rev.\ B {\bf 88}, 035119 (2013)};
M.\ van Veenendaal,
\href{http://dx.doi.org/10.1103/PhysRevB.87.235118}
{Phys.\ Rev.\ B {\bf 87}, 235118 (2013)}.

\bibitem{lazarovits2010}
B.\ Lazarovits, K.\ Kim, K.\ Haule and G.\ Kotliar,
\href{http://dx.doi.org/10.1103/PhysRevB.81.115117}
{Phys.\ Rev.\ B {\bf 81}, 115117 (2010)}.

\bibitem{goodenough1973}
J.\ B.\ Goodenough and H.\ Y.-P.\ Hong,
\href{http://dx.doi.org/10.1103/PhysRevB.8.1323}
{Phys.\ Rev.\ B {\bf 8}, 1323 (1973)}.

\end{thebibliography}
\end{document}